# JoKER: Trusted Detection of Kernel Rootkits in Android Devices via JTAG Interface

Mordechai Guri, Yuri Poliak
Department of Information Systems Engineering
Ben-Gurion University
Beer-Sheva, Israel
{gurim,yuripo}@post.bgu.ac.il

Bracha Shapira, Yuval Elovici
Department of Information Systems Engineering
Telekom Innovation Laboratories at Ben-Gurion University
Beer-Sheva, Israel
{bshapira,elovici}@bgu.ac.il

*Abstract*— Smartphones and tablets have become prime targets for malware, due to the valuable private and corporate information they hold. While Anti-Virus (AV) program may successfully detect malicious applications (apps), they remain ineffective against low-level rootkits that evade detection mechanisms by masking their own presence. Furthermore, any detection mechanism run on the same physical device as the monitored OS can be compromised via application, kernel or boot-loader vulnerabilities. Consequentially, trusted detection of kernel rootkits in mobile devices is a challenging task in practice.

In this paper we present '*JoKER*' - a system which aims at detecting rootkits in the Android kernel by utilizing the hardware's Joint Test Action Group (JTAG) interface for trusted memory forensics. Our framework consists of components that extract areas of a kernel's memory and reconstruct it for further analysis. We present the overall architecture along with its implementation, and demonstrate that the system can successfully detect the presence of stealthy rootkits in the kernel. The results show that although JTAG's main purpose is system testing, it can also be used for malware detection where traditional methods fail.

Keywords— Android; Forensics; JTAG, Rootkits; Security; Trusted Detection;

## I. INTRODUCTION

Over the last few years, mobile devices have emerged as a preferred target for cyber criminals. This trend is fueled by the lucrative private and organizational information stored on those devices. Android is by far the most popular mobile Operating System (OS); its numerous vulnerabilities, coupled with the ease of distributing malicious code through its flexible app market, have turned this OS into the attackers' favorite target [1]. For example, the Droid Dream attack [2] distributed through legitimate applications on the Android market infected about 50,000 mobile devices in a course of few days. More recently, an Android 'bootkit', i.e. a rootkit that modifies the device's boot partition and booting script (codenamed 'Oldboot') has infected more than 500,000 mobile devices in China alone, within a period of six months [3] .

### A. Kernel Rootkits

Mobile and desktop malware can operate in user space or kernel space. The user space malware can modify and inject code only into the memory areas allocated to apps and user processes. The kernel space malware can manipulate objects that reside in the entire memory area of the OS. Although sophisticated Mandatory Access Control (MAC) mechanisms such as SElinux [4] are integrated into current versions of Android, malware developers still manage to run their code in the kernel [5] [6] [3]. Rootkits are kernel space malware that use illicitly granted exclusive permissions to hide the malware's existence from detection systems, by manipulating the kernel's internal data structures [7]. A malicious code that has penetrated into the memory of the kernel can neutralize any security tool running in the OS. For instance, if an process sends a request to the kernel asking for the list of files in a specific directory there is no guarantee for the integrity of the returned list. Consequently, in order to detect presence rootkits, a *trusted* snapshot of the kernel memory has to be obtained [8].

### B. The Proposed System

In this paper we present *JoKER (JTAG observe Kernel)*, a system that utilizes the JTAG hardware interface of the mobile device in order to obtain trusted snapshot of the device memory for detection of kernel rootkits. The JTAG standard [9] was developed to assist with system testing and debugging the circuit board after manufacturing. JTAG's connectors are installed on the Printed Circuit Board (PCB) of modern mobile devices such as smartphones and tablets. Our detection system uses two important debugging features of JTAG:

1. The ability to *halt* the system instantly by sending special instruction to the main processor.

2. The ability to *access* the content of the device's volatile memory (RAM) while it is being halted. The overall system does not run on the mobile device and therefore can securely read the kernel's memory areas in a trusted manner.

Once the kernel memory is extracted, it is passed through an array of programmed scripts. Each script reconstructs specific data-structures in the kernel and analyzes them for traces of suspicious modifications. We present the system architecture and discuss the implementation in details. Our evaluation shows that JoKER can successfully detect maliciously modified objects at the Android kernel in a trusted manner.

*C. Method Limitation*

Using the JTAG interface requires physical connection to the JTAG port which is placed on the smartphone's main board. Compared to software-based methods, this approach may appear rather awkward. However, having external memory acquisition capabilities (from outside of the device) offers the advantage of trusted memory inspection. Accordingly, our proposed system aims at finding stealthy and sophisticated rootkits where other detection methods, running within the device, cannot be trusted.

*D. Our Contribution*

JTAG, as a general forensic tool for embedded devices and Android systems was mentioned in prior work [10] [11]. This paper introduces several contributions and advantages over prior related work in the field.

- First, we are the first to propose an automated system focused on detecting kernel rootkits for the Android OS and ARM architecture, by utilizing JTAG based memory forensics. Our method is trusted since it is external, hardware-based, and transparent to the malicious code, hence cannot be subverted. We present the overall architecture and detailed working implementation of the detection system, at both hardware and software levels.
- Second, we discuss five rootkit mechanisms for the Android kernel and show how they can be identified by our system.
- Third, we introduce a new method for detecting hidden processes by analyzing the Android kernel cache mechanism.
- Forth, we show how to overcome several challenges involved with our low-level examination of the system. Those challenges include translating between physical and virtual memory addresses, along with resolving notorious kernel synchronization issues.

## II. RELATED WORK

While existing mobile antivirus apps may detect user space malware, they are in general not effective for the detection of kernel level rootkits [5] [12]. Tools such as the Linux Memory Extractor (LiME) [13], and DMD [14] are helping for acquisition and analysis of volatile memory in Android. However, since these tools operate from within the OS they can be subverted by a rootkit, hence cannot be considered to be trusted. Android kernel securing and hardening [15] [16] [17] has been also proposed for defending the kernel memory space against rootkits. Hypervisors [18] [5] [19] and Trusted Platforms Module (TPM) [20] [21], had been researched to provide trusted point of acquisition for kernel space memory.

More recently sun et al presented TrustDump, a hardware-assisted system for reliable memory acquisition on smartphones using ARM TrustZone support [8]. The mechanism employed by TrustDump is running in the TrustZone's secure domain ensure an isolation between the OS and the memory acquisition tool. However, as long as the security mechanism runs on the same physical device as the monitored OS, it can be compromised via runtime or boot-loader vulnerabilities [22] [6]. JTAG was discussed as a tool for forensic imaging of embedded system [11] and in context of Android devices [10] in a general manner. Table 1 shows the different layers used for malware detection and memory acquisition in mobile devices.

**Table 1: Different layers of malware detection and memory acquisition mechanisms for mobile devices**

| Approach | Implementation | Run In-device | Compromise Methods |
|---|---|---|---|
| **Kernel** | [15] [16] | Yes | [3] [12] [19] |
| **Hypervisor** | [18] [23] | Yes | [22] [6] |
| **TrustZone** | TrustDump [8] | Yes | [22] [6] |
| **JTAG** | JoKER | No | |

## III. SYSTEM DESIGN

The JoKER system consists of four components, as depicted in Figure 1; (A) the mobile device (B) the JTAG controller (C) a program that extracts the kernel raw memory from the device and manages its analysis, and (D) set of scripts to analyze and detect rootkits at selected areas of the kernel. The mobile device is an Android device that is scanned for the presence of rootkits. This device should have a JTAG port with a compatible soldered connector so that it can be connected to a JTAG controller. The JTAG controller is the hardware component that can communicate with the CPU and the memory controller on the target device through the On Chip Debug (OCD) connectors. The memory extraction program receives the raw content of the device's RAM by communicating with the OCD. Finally, the set of scripts reconstructs and analyzes the kernel memory.

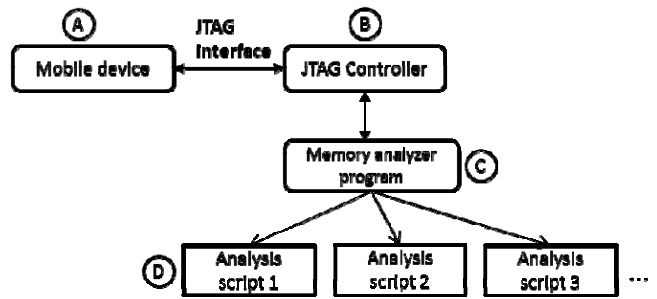

**Figure 1: Schematic layout of the system's components**

The detection process consists of three main phases: (1) halting the processor of the target device, (2) extracting kernel's data structures from the RAM, and (3) applying a forensic analysis algorithm to find rootkit footprints in the extracted binaries. These steps are described in the following subsections.

### A. System Halting

JTAG can halt the core of the mobile device by sending halt command to the OCD [24]. We use this functionality at an initial stage to ensure that no code is executed on the device. This fact plays a major role in the detection mechanism's design since the suspicion that the system is being monitored can prompt a running malware to mask its presence. Halting the processor in a single command ensures that a malware cannot prepare its masking before the halt.

### B. Extracting Kernel Memory

The second phase involves extraction of kernel related memory areas for further analysis in a separate computer. Modern JTAG interfaces offer rich debugging functionality such as direct Read and Write access to the RAM and Flash memory [24]. We have used JTAG's commands to extract raw memory from the RAM of the device while it is halted. The decision about the memory regions to extract is based on the specific analysis techniques. For demonstration of the system, we used techniques adapted from studies related to Linux-based rootkits [25]. Rootkits attack various data structures on Linux systems, primarily the system call table, exception vector table and the kernel's processes list. We therefore focus on extracting the related memory regions for further analysis.

### C. Reconstruction and Analysis

During the third and final phase, the detection system applies analysis algorithms to the extracted raw memory. This process involves scanning for suspicious modifications of memory regions. The scripts check the integrity of the system call table, the exception vector table, and the software interrupt handler. Since these objects shouldn't be modified on a regular Android system, we validate their integrity compared to a clean Android system.

Another script detects stealthy processes which are hidden from the kernel's processes list. Unlike the system call table, the exception vector table, and the software interrupt handler, the processes list is a *dynamic* kernel object which is changed frequently. Detecting hidden rootkit processes is challenging since rootkits typically remove their entry from this list in order to evade detection. To that end, our system analyzes the kernel's cache which is responsible for maintaining pools of the OS internal objects. We have applied cross-view methodology by comparing the objects in the kernel's processes list to a baseline that consists of active processes reconstructed from the protected cache pools. A difference between the two views indicates the presence of hidden processes. This method may reveal the presence of a rootkit and can also pinpoint the processes that the rootkit has tried to hide.

## IV. IMPLEMENTATION

We have implemented the JoKER framework according to the design outlined in section II. We used the RIFF Box JTAG controller [26] to communicate with a JTAG capable Android device. The tests described in this paper were conducted on Samsung Galaxy (S2 & S4) mobile phones with JTAG interface. On the software side, the JTAG controller (Figure 1, B) extracts relevant memory regions from the device (Figure 1, A) via a set of PRACTICE scripts. PRACTICE [27] is a scripting language which operates on the Lauterbach TRACE32 microprocessor development tool and its related product line. Those tools are aimed at providing easy programming access to on-chip debug systems. TRACE32 supports communication with the JTAG interface among other interfaces. The memory analyzer program (Figure 1, C) receives the raw memory data and feeds it to a set of Python scripts (Figure 1, D). Each script receives the memory contents as an array of bytes, performs its own forensic analysis and returns the results. All logs are saved by the memory analyzer program and the final results are presented to the user.

### A. System setup

Figure 2 presents the system setup as constructed and installed in our lab.

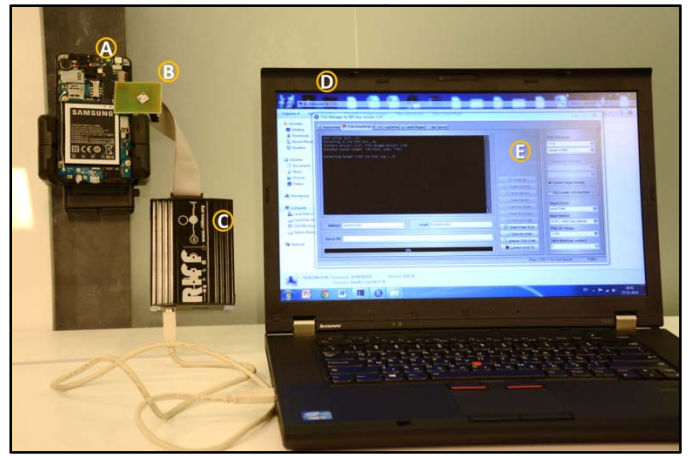

**Figure 2: The JoKER system, as constructed and installed in the lab**

The setup follows the outline discussed earlier and consists of the RIFF Box JTAG controller (Figure 2, C) which supports communication with a variety of mobile and embedded devices on the market. The JTAG controller is connected to the JTAG port on the target device (Figure 2, A), through a flat cable with a connector to the JTAG port on the device (Figure 2, B). The other end of the JTAG box is connected through a USB cable to a computer that hosts the controller program (Figure 2, D). The laptop computer also hosts the PRACTICE scripts that are responsible for the memory extraction and the Python scripts that are responsible for the memory analysis (Figure 2, E).

*B. Memory Analysis*

JoKER is a generic framework which can be enriched with wide range of detection and analysis scripts. For the system testing we implemented five scripts, each targeting a different type of rootkit technique. The scripts include: (1) system call table integrity checks, (2) exception vector table (EVT) integrity checks, (3) two types of software interrupt handler (SWI handler) integrity checks and (4) revealing hidden process by analyzing the kernel's cache. To the best of our knowledge, the former method is new and introduced for the first time in this paper. The analyzed kernel structures are presented in Table 2. For clarity, a flow of a system call in the Android kernel along with the relevant tables is outlined in Appendix A.

**Table 2: The list of Android kernel object analyzed by our JoKER implementation**

| Structure | Description |
|---|---|
| **System Call Table** | A static structure which contains pointers to low level system functions |
| **Exception Vector Table (EVT)** | A static structure which contains pointers to exceptions and interrupt handlers |
| **Software Interrupt Handler (SWI)** | A static structure which contains pointers to interrupt handlers |
| **kmem_cache structure** | Dynamic structure which contain information on kernel cached objects |

Prior to the system operation, the analysis scripts are initialized with the physical address of the objects within the kernel memory in the specific version of examined Android. These parameters can be extracted from the kernel's symbol list located at /proc/kallsyms. Note that these parameters can be retrieved from any clean device having the same version of the kernel. We have developed a Loadable Kernel Module (LKM) which is executed on a clean (downloaded from the official website) Android device with an identical version of the Kernel and report the parameters' values back to the system.

*1) Physical to virtual memory translation*

Since JTAG refers to memory in physical addresses, we had to translate between the virtual addresses (OS view) and the physical addresses (JTAG view). As the input addresses are part of the kernel space, they can be calculated from the virtual address by subtracting a fixed offset. An exceptional case is the address of the EVT, since on ARM based architectures the virtual address of the EVT must be either 0x00000000 or 0xffff0000. To calculate the physical address of the EVT we used ARM's assembly instructions which translate virtual to physical address by traversing the page tables in our LKMs. Note that on Android distributions that disable the LKM mechanism and omit the kernel's symbol list, it is still possible to extract the initializing parameters by using the Runtime Kernel Patch (RKP) strategy for accessing the kernel space memory as has been demonstrated in [7].

*C. Detection Scenarios*

Some detection techniques involve checking the integrity of various structures of the kernel. This scenario is relevant when performing a 'before and after' forensic examination, for example, when examine an application that may bring a malicious payload into the device. In such cases the forensic analyst will have to examine the system at two points: before installing the application (a 'clean' snapshot) and after installation.

*D. Dectection Flow*

The flow chart in Figure 3 outlines the process of rootkit detection from the time that the initializing parameters have been set, the JTAG controller has been connected to the target device, and the communication with the control software has been initialized. We assume that a clean snapshot of the kernel's memory has been taken previously from a device with a clean kernel version (E.g., downloaded from official website).

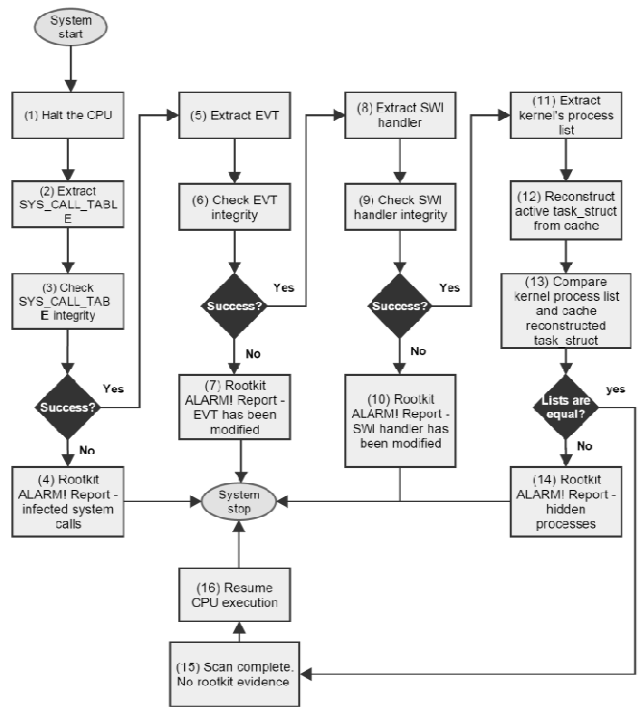

**Figure 3: Outline of the detection flow in our implementation of JoKER**

The main steps of the detection algorithm are as follows:

**Steps (1-4):** Halting the CPU of target device and validating the integrity of the current system call table against a clean version of this table. In cases of inconsistency, a rootkit alert is triggered. **Steps (5-7):** Validating the integrity of the current EVT against a clean version of the EVT. In cases of inconsistency, a rootkit alert is triggered. **Steps (8-10):** Validating the integrity of the SWI handler pointer against clean version of the SWI handler and validating the integrity of the SWI handler code. In a case of inconsistency, a rootkit alert is triggered. **Step (11):** Extracting the kernel's processes list, by parsing each task_struct node in the list starting at the INIT process. **Step (12):** Reconstructing the list of task_struct that appear in the cache mechanism of the kernel. **Step (13):** comparison between the list of task_struct which was extracted from the kernel's list, and those extracted from the cache. **Step (14):** If the kernel's process list and the cache differ by the task_structs, a hidden process is found, and rootkit alert is triggered.

## V. EVALUATION

We evaluate the detection system by testing it against five types of suspicious kernel modification code. To that end we implemented five kernel modules which perform the malicious operations. The reason for using self-constructed rootkits rather than original ones is due to the fact that no samples of rootkits for current mobile phones have been released to the research community (source or binary). Interestingly, although kernel rootkits have been widely researched in the context of desktop operating systems, there are no documented samples of rootkits for recent versions of Android. In addition, rootkits which target the desktop version of Linux kernel cannot be installed on the Android kernel. This is due to differences in the kernel architecture between the OSs, and the modified versions of LIBC in the Android OS. We evaluate the system with Android kernel 2.6.35 and 3.4.0 installed on Samsung Galaxy S2 and Galaxy S4, respectively.

### A. Kernel Rookits

The rootkit mechanisms have been implemented in the form of LKMs. Each of the five rootkits (Samples 1-5) exposes a different malicious functionality. Sample 1 modifies the address of four system calls in the system call table. Sample 2 modifies an indirect pointer of the SWI handler which is stored in the instruction at offset 0x8 in the EVT. Sample 3 modifies the address of the SWI handler which is stored in the EVT. Sample 4 modifies the offset of the system call table which is stored in the SWI handler routine. Sample 5 hides a process by removing it from the kernel's processes list. We use self-implemented rootkits is mainly due to the fact that only a few modern samples of mobile rootkits have been published. Note that self-implemented rootkits as an evaluation method is has been used in previous studies in the field [18] [12].

Figure 4: System-call table, before (upper) and after (lower) the rootkit operation

### B. Syscall Table Hooking

The first rootkit was implemented as a kernel module (syscallTableHook.ko[1]) which modifies the address of four system calls addresses in the system-call table; read(), write(), open(), and close(). We choose four basic system functions that can be used maliciously in order to intercept file system, sensors and network access operations. The experiment starts by executing a PRACTICE script to get a binary snapshot of the kernel's system call table before and after the execution of the rootkit.

In Figure 4 we see the two snapshots of the system call table in a binary form of hex editor viewer. As can be seen, four modified addresses in the table have been detected. The original addresses of the system calls are marked in blue while the modified addresses are marked in red. In the next step, the script checks which system calls have been changed. This is achieved by parsing the header file (unistd.h) from the source tree of the Android kernel. This file contains the order and the names of system call functions in the table. Next, the Python script receives the two snapshots of the system-calls table and the list of functions from the kernel and returns the names of functions that were modified. The output of the system is shown in Figure 5.

Figure 5: Output of the analysis script in which the modified system are detected and identified, along with their addresses and names

---

[1] Source of the five kernel modules will be published online

## C. Exception Vector Table (EVT) hooking

In the ARM architecture, each exception or interrupt is branched to the Exception Vector Table (EVT). This table is a central component of the OS and is naturally a target of different hooking techniques. When a software interrupt happens in the system, the processor loads the instruction at offset 0x8 in the EVT to the instruction register for execution (Figure 6). The instruction that will execute in this case is ldr pc, [pc, #1040]. This instruction loads the program counter with the address of the software interrupt handler address that resides in the offset 1040 (0x420) relative to the current program counter

```
[000] ffff0000: ef9f0000 [Reset]        ; svc 0x9f0000 branch code array
[004] ffff0004: ea0000dd [Undef]        ; b    0x380
[000] ffff0000: e59ff410 [Swi]          ; ldr pc, [pc, #1040] ; 0x420
[00c] ffff000c: ea0000bb [Abort perfetch] ; b    0x300
[010] ffff0010: ea00009a [Abort-data]   ; b    0x280
...
[420] ffff0420: c003df40 |vector_swi|
```

**Figure 6: A snapshot of the EVT in the kernel**

The second rootkit was implemented as a kernel module (HookBranchInstruction.ko) which modifies the exception vector table. Our implementation technique is similar to [7], applying two types of modifications to the exception vector table. First, it copies the address of a new SWI handler to the memory at offset 0x424 in the exception vector table. Second, the rootkit changes the instruction at offset 0x8 to load the address at offset 0x424 (the new handler) to the table instead of the original address. This technique allows the attacker to hook the SWI handler and intercept interrupts and exceptions in the system. Our system extracts the kernel's memory before and after the rootkits installation, by using PRACTICE script. A Python script reconstructs and compares the two views.

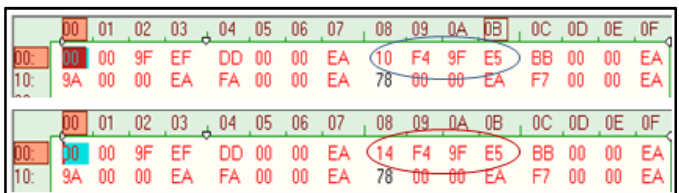

**Figure 7: The EVT, before (upper) and after (lower) the rootkit's modification**

As can be seen in Figure 7, the instruction at offset 0x8 of the exception vector table has been changed from 0xe59ff410 to 0xe59ff414. This modification causes the processor to load the address that resides at offset 0x424 of the table instead of the address at offset 0x420. The difference between the two EVTs is identified and reported to the system as a rootkit alert.

## D. Hooking the Address of SWI Handler Routine

Another hooking approach is modifying the SWI handler routine. Basically a rootkit injects the address of its own handler function. By intercepting all interrupts and exception calls, the rootkit can perform malicious operations in a hidden manner. In our example the rootkit (exvHookSwiHandlerAddress.ko) copies the binary content of the original SWI handler to another address in the kernel space, modifies the binary code of the handler and then puts the address of the new handler at offset 0x420 of the exception vector table. Our system extracts the kernel's memory before and after the rootkits installation, by using a PRACTICE script. Then, a Python script reconstructs and compares the two views.

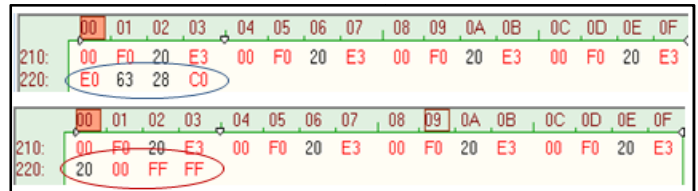

**Figure 8: The address of SWI handler, before (upper) and after (lower) the rootkit's modification**

As can be seen in Figure 8, the address of the SWI handler has been changed at offset 0x220 in the second part of the exception vector table which is offset 0x420 from the base of the table. This difference indicates that a malicious modification has occurred. The event is reported to the system as a rootkit alert.

## E. Hooking the Code of SWI Handler Routine

The last hooking technique involves hooking the binary code of the software interrupts routine itself.

```
000000c0 <vector_swi>:
...
100: e1a096ad mov    r9, sp, lsr #13 ; get_thread_info tsk
104: e1a09689 mov    r9, r9, lsl #13
108: e28f8094 add    r8, pc, #148    ; load syscall table pointer
10c: e599c000 ldr    ip, [r9]        ; check for syscall tracing
```

**Figure 9: Part of the SWI hander routine code in the kernel memory**

Figure 9 shows the part of the SWI handler that locates the address of the system call table with an offset relative to the current Program Counter. The system call table itself is located after the code of the handler. By manipulating the marked instruction, a rootkit can direct any software interrupt to its own system call functions. Our implemented rootkit (hookSysCallTableAddressInSwiHandler.ko) iterates over the entries of the instruction which loads the system-call table pointer. Next, the instruction is replaced with a new instruction - ldr r8, [pc, #offset], where #offset is the relative offset of our system-call table.

Figure 10 depicts modifications of the SWI handler routine identified by our detection system. A NOP instruction (0xe320f000) has been changed to the address of the malicious system call table (0xc02864c8). The instruction that loads the address of the system call table into register r8 has been changed from add r8, pc, 0x98 to ldr r8, [pc, 0x80]. The PRACTICE script generates snapshots of the SWI handler, the

Python script compares them, and a rootkit alert is triggered when a relevant modification is detected.

**Figure 10: SWI code in memory, before (upper) and after (lower) the rootkit operation**

*F. DKOM (Direct Kernel Object Manipulation)*

Direct Kernel Object Manipulation (KDOM) is a technique used by a rootkit in order to hide itself from the OS layer. By directly accessing the data structures in the kernel, a rootkit can hide resources such as processes and threads descriptors, network connections and other objects in the memory. To examine the effectiveness of our system against DKOM, we implemented a rootkit (dkomRootkit.ko) which manipulates the linked list of the kernel's structures representing the list of processes and threads (task_structs). We executed a process on the device which simulates the malicious program (MalApp) that the rootkit intends to hide. The program itself is executed as a user-level process. Our rootkit scans the linked list of the kernel's task_structs, searching for a task with the name "MalApp", and removes it. Note that although the process is removed from the link list, it still exists in the scheduler's internal list; hence its execution is not terminated. To detect the hidden process, we developed a new cross-view strategy which uses the kernel's cache pool. The kernel cache contains the cached version of the task_struct while it is in use, or shortly after termination for reuse. Rookits typically do not interfere with the cache pool as it is an internally managed memory region. We used this fact to make analysis of the cache pool and identify traces of hidden processes. Our script reconstructs the processes list from the kernel's processes list and from the cache. The results of the comparison (Figure 11) show that all the tasks appearing in the linked list also appeared in the cache, but there is a task_struct that appears in the cache but is not part of the linked list of processes descriptors.

For the interested reader we point that in most Android distributions the cache mechanism does not have pointers to all slabs [28] that contain task_structs. Therefore, we obtained the slab addresses in the following manner: We traversed each Page Frame Number (PFN), translated it into a physical struct page address, and then checked whether the struct page represent a slab with task_struct objects. From each matching slab, we extracted all task_structs.

**Figure 11: Results of the comparisons (cross-view) between the kernel's processes list and the process list reconstructed from the cache**

*1) Kernel Consistency*

When evaluating the cross-view detection approach, we noticed that when the list of task_structs is extracted from the cache on the clean system, some of the structures might not appear in the kernel's process list. Although rare, this behavior should be understood and eliminated when dealing with clean systems. We found that the reason for this exceptional behavior is the way that the JTAG box communicates with the device. When our system starts executing any of the PRACTICE scripts, the processor of the target device is instructed to halt immediately. The problem is that when halting occurs, the kernel of the device is, very briefly in an unstable state. The cache mechanism reuses the objects of task_struct. Thus when a process ends its execution, the kernel should unlink it from its list of processes and only then mark it as an unused object in the cache. This process of object reuse is not an atomic operation, and the halting of the core of the system take place in the middle of the unlinking operation. This momentary unstable state causes some active processes to appear as if absent from the kernel's processes list. To distinguish between malicious processes (intentionally absent from the list) and "dummy" processes (absent because of the inconsistency), we have analyzed the task_structs of these "dummy" processes. In so doing, we have determined that the fields of pid, comm, state and flags in the task_structs can serve to indicate whether it is being halted. In some of these objects the value of the pid was 0 but the name of the process (comm) was not "swapper". Obviously, such an object cannot represent a runnable process as the only process in the system with pid 0 can be the swapper. Other active objects had a negative value in their state field. This field contains information about the runnable state of the process, and a negative value represents a non-runnable state. The last indicator is the flags field which contains information about the state of the process. The value of this field is a bitwise-OR of all the characteristics that represent the state of the process at the moment. If the least significant bits equal 2, then the process is in a shutting down mode. Since the kernel's non-consistent task_structs can be filtered by the indicators listed above, we redesigned the detection system to filter these objects before comparing task_structs in the cache and in the linked list. We executed our redesigned detection mechanism on a typical clean system and validate that it does not issue false alarms as a side effect of the kernel cache behavior.

VI. CONCLUSION

In this paper we present JoKER, a framework which utilizes the hardware's JTAG interface for trusted memory forensics. Our system demonstrates how kernel level rootkits in the Android OS can be detected in an automated manner by employing various memory forensic techniques. Unlike conventional methods, our method is trusted since it is external, hardware-based, and undetectable by the malicious code within the device.

JoKER framework extracts areas of the kernel's memory, reconstructs them for further analysis, and raises a rootkit alert when positive evidence is encountered. We present the overall layout of the framework, along with its detailed implementation. Our system is evaluated under several attack patterns, demonstrating that it can successfully detect crafty kernel mode rootkits, whether persistent or non-persistent. We present the implementation of five types of rootkits, used for evaluating our system, and show how our system detects them. In particular, a new method is introduced for detecting hidden processes by analyzing the Android kernel cache data structure. We also discuss some technological challenges involved with our method, such as translation between physical and virtual memory addresses and resolving kernel synchronization issues. The detection system demonstrates the cross-view paradigm in which the inspected system is examined at multiple levels in order to expose contradicting traces suggesting the presence of a rootkit, and eliminate false alarms. Note that, although the original purpose of JTAG is system testing and verification, in this paper we show that it can also be used for low level malware detection. We believe that our current experimental system can serve as a platform or prototype for future research concerning trusted detection of mobile devices rootkits and similar kernel-level malware.

APPENDIX A

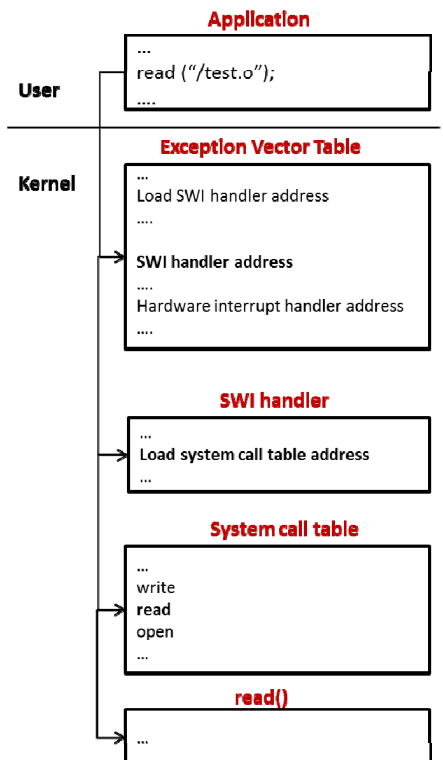

**Figure 12: The flow of 'read' system call in Android OS from the application level to the kernel level**